\documentclass[%
 reprint,
 amsmath,amssymb,aps,
 prl,
]{revtex4-2}

\usepackage{graphicx}% Include figure files
\usepackage{dcolumn}% Align table columns on decimal point
\usepackage{bm}% bold math
\usepackage{xcolor}
\usepackage{hyperref}
\hypersetup{colorlinks=true,linkcolor=blue,citecolor=blue, urlcolor=blue }
\begin{document}

\preprint{APS/123-QED}

\title{Space-Time Electron Wave Packets with Tunable Skyrmion Textures}% Force line breaks with \\

\author{Sharoz Schezwen}
 \affiliation{School of Electrical and Electronics Engineering, Nanyang Technological University, Singapore}%Lines break automatically or can be forced with \\
\author{Yijie Shen}%
\email{yijie.shen@ntu.edu.sg}
 \affiliation{Centre of Disruptive Photonics Technologies, School of Physical and Mathematical Sciences, Singapore}
 \affiliation{School of Electrical and Electronics Engineering, Nanyang Technological University, Singapore}

\date{\today}

\begin{abstract}

Space-Time Wave Packets (STWPs) have enabled numerous nontrivial properties, such as tunable group velocities, nondiffraction and nonlocal spin textures in optical fields. However, this has not been properly studied in matter waves. Here, we investigate free-electron STWPs, which are propagation-invariant and group-velocity-tunable matter wave packets, constructed by constraining the spatiotemporal spectrum of the electron wavefunction in momentum space. Due to the unique dispersion of free-electrons, the electron STWP can carry two different topologies, and host both meron and skyrmion topological textures in momentum space. We demonstrate the topological phase transitions and tunable skyrmion numbers by spectral geometry tuning. We also show the topologically robust propagation of the electron pulses and corresponding real-space skyrmionic expectation value textures encoded within each single electron wavefunction. Our results open a platform for topological free-electron wavepackets uses in sensing and information transfer.
\end{abstract}

\maketitle

\textit{Introduction - }Structured free-electron waves have emerged as a reliable and powerful platform to control quantum matter waves, enabling the creation of electron vortex beams carrying angular momentum \cite{McMorran_Agrawal_Anderson_Herzing_Lezec_McClelland_Unguris_2011,Lloyd_Babiker_Thirunavukkarasu_Yuan_2017,Fang_Kuttruff_Nabben_Baum_2024,Fang_Kuttruff_Baum_2026}, to shaped electron wavepackets used for microscopy and spectroscopy \cite{Madan_Leccese_Mazur_Barantani_LaGrange_Sapozhnik_Tengdin_Gargiulo_Rotunno_Olaya_etal._2022,Reinhardt_Kaminer_2020,Greig_Khurgin_Elezzabi_2020}. Yet single free-electron wavepackets remain fundamentally limited by issues such as the diffractive spreading and dispersion, and temporal walk off that limits the interaction lengths in free-electron-photon interactions; persistent bottlenecks in electron optics applications \cite{Gliserin_Walbran_Krausz_Baum_2015, Arrowsmith_Miller_Ambat_Bahk_Begishev_Bromage_Bucht_Dauphin_Dorrer_Jeon_et_al._2026, Pennacchio_Vanacore_Mancini_Oppermann_Jayaraman_Musumeci_Baum_Carbone_2017}. Beyond this, a free-electron has an intrinsic spin degree of freedom, fundamentally limiting the possibility of creating spin textures for single electrons \cite{Wang_Li_2011}. Such spin textures may provide fast spin filters for measuring material spin textures from the Hund's exchange term in the interaction Hamiltonian \cite{Denisov_Rozhansky_Averkiev_Lahderanta_2016,Swain_Mandal_Satpathy_Nanda_2026} (see Supplemental Material F), without the need for complex vector reconstruction or sample tilting.

\begin{figure}
    \centering
    \includegraphics[width=0.9\linewidth, height=5.5cm]{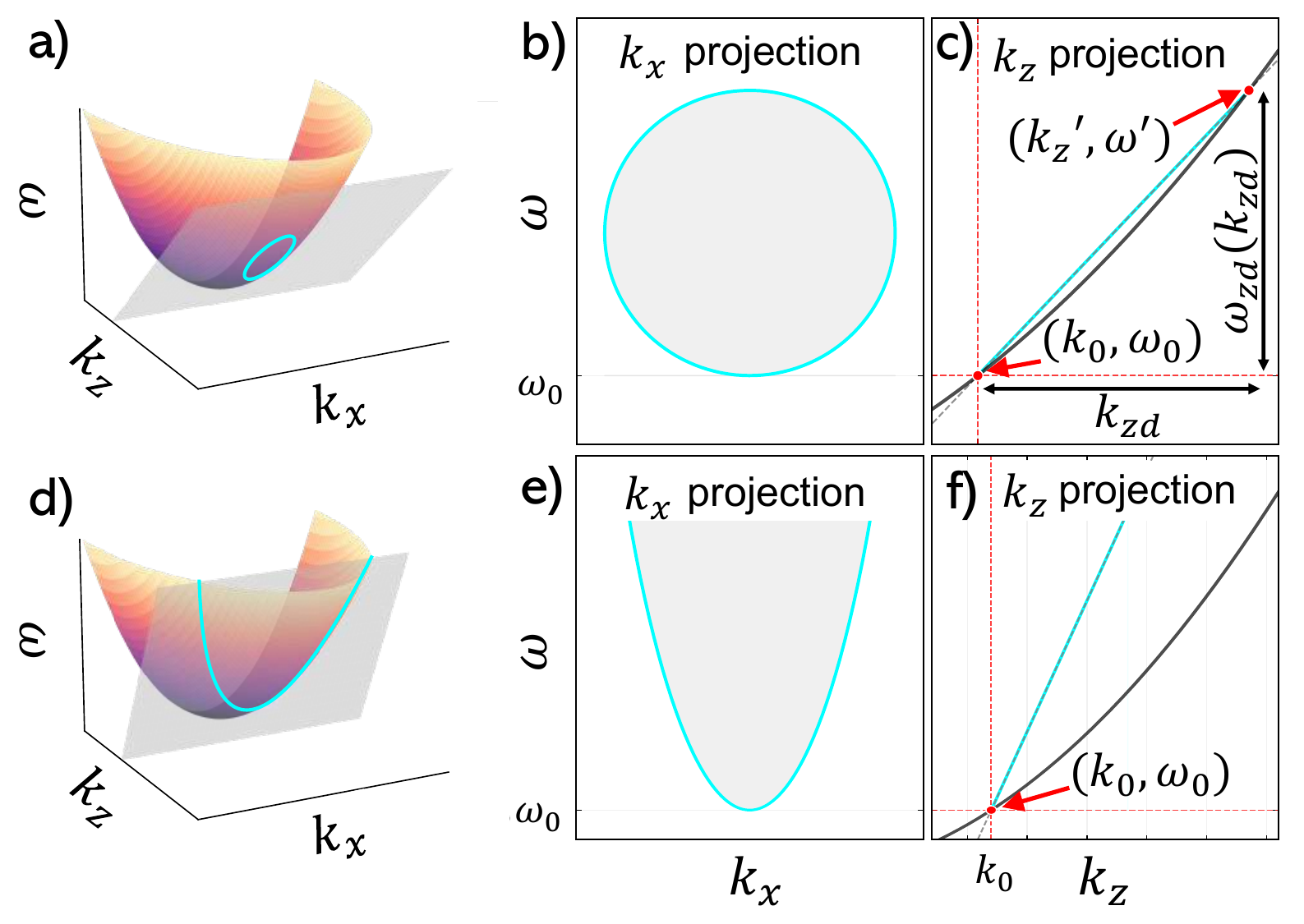} %fig1.jpg
    \caption{a) The construction of momentum space skyrmion uses a closed spectral loop in b) $k_x-\omega$ space. c) The intersection of the spectral plane and the dispersion surface gives the lowest point ($k_0,\omega_0$) called the carrier, and the point ($k_z',\omega'$). The difference between them is $k_{zd}$. When $\omega'$ is the maximum allowed constituent electron plane wave energy, we write the point as ($k_{zmax},\omega_{max}$) in text.  d-f) Merons can be obtained from an intersection that produces a parabolic projection in the $k_x-\omega$ plane instead.}
    \label{fig:Fig1}
\end{figure}

Recently, Space-Time Wave Packets (STWPs) are an emerging class of structured waves with a correlated space and time degrees of freedom, which gives the waves a well defined and tunable group velocity and great resistance against ordinary diffraction and dispersion \cite{Hall_Abouraddy_2025,Kim_Guo_Catrysse_Fan_2025,Kondakci_Abouraddy_2016,Kondakci_Abouraddy_2019,Romer_Hall_Abouraddy_2025,Su_Zou_Wang_Yessenov_Zhou_Song_Ko_Zeng_Alhaddad_Ramakrishnan_et,Wang_Guo_Fan_2025,Yessenov_Abouraddy_2020,Yessenov_Hall_Schepler_Abouraddy_2022,Yessenov_Free_Chen_Johnson_Lavery_Alonso_Abouraddy_2022,Kondakci_Abouraddy_2017,Zhan_2024,Che_Liu_Ye_Shi_Chan_Zi_2024,Liu_Ge_Xu_Sun_Liu_Lu_Chen_2025,Pitanti_Ashurbekov_dePedro-Embid_Msall_Santos_2025,Singh_Ronning_Liu_Angheluta_Concha_Bandi_2026,Smirnova_Nori_Bliokh_2024,Zhang_Sun_Huang_Wu_Yang_Bliokh_Ruan_2023}. An optical STWP was also shown to carry a nondiffracting meron, by using a metasurface to imprint a polarization texture in momentum space \cite{Yessenov_Dorrah_Guo_Hall_Park_Free_Johnson_Capasso_Fan_Abouraddy_2025}, while related momentum space merons have also been realized through photonic crystal slabs \cite{Guo_Xiao_Guo_Yuan_Fan_2020, Guo_Xiao_Orenstein_Fan_2021}. The merons are a half skyrmion, topologically nontrivial spin textures originally introduced in nuclear physics, which have since been obtained in other fields \cite{SKYRME1962556,Shen_Wang_Fan_2025,Psaroudaki_Panagopoulos_2021,Shen_Wang_Fan_2025,Shen_Zhang_Shi_Du_Yuan_Zayats_2024,Teng_Liu_Zhang_Fan_Chen_Cao_Zhong_Lei_Zhan_2025,Niu_Feng_Zhang_Yu_Wang_Lang_Xu_Chen_Ma_Qiu_et,Shen_Hou_Papasimakis_Zheludev_2021,Wang_Ying_Shi_Wang_Wang_Liang_Shen_2025,Zdagkas_McDonnell_Deng_Shen_Li_Ellenbogen_Papasimakis_Zheludev_2022,Shen_Papasimakis_Zheludev_2024,Wang_Bao_Hu_Shi_Wang_Zheludev_Shen_2024}. The key feature of a skyrmionic texture is the topological robustness, making it attractive for robust transport and information encoding \cite{Wang_Zhao_Ma_Cai_Zhang_Shang_Zhang_Qin_Pong_Marozsak_et,Guo_Peters_Mata-Cervera_Vetlugin_Guo_Zhang_Forbes_Shen_2026, Zhang_Wang_Zhang_Zhao_Ma_Liu_Pong_Cai_He_2026}. In optical STWP, the fundamental structure of the light cone makes the meron the only naturally available topological structure. However, non-relativistic free-electrons obey a quadratic dispersion relation ($\omega = \hbar k^2 /2m$), see Fig.~\ref{fig:Fig1}a the dispersion surface. This allows a free-electron STWP to be nondiffracting and nondispersive, have tunable group velocities, and to host full or half skyrmionic topologies in momentum space and the corresponding realspace wavefunction. Such realization gives an entirely new degree of freedom for free-electrons.

In this paper, we show this construction of nonrelativistic free-electron STWP by engineering the spatiotemporal spectrum on the quadratic dispersion relation. We show despite being propagation invariant, such construction also supports momentum space skyrmions and merons, and we further derive the analytical topological phase transition between them. In realspace, we show the existence of propagation invariant skyrmion and meron textures in the local spin expectation value that remains robust upon extended propagation distances. This establishes a route toward topologically structured electron wavepackets whose real-space topologies are transported unchanged in free space, for use in sensing and information transfer.

\textit{Results - }To construct an electron STWP, the free-electron dispersion relation is given by the solution to the Schrodinger equation in a closed system, written as $\omega = \hbar k^2/2m$, where $k^2$ and $m$ is the momentum and the mass of the elementary electron. Throughout this paper, both $\hbar$ and $m$ is taken to be 1 for the analytical calculations. A spectral constraint in the form of a plane ($\omega = \omega_0+\tan\theta (k_z - k_{0})$) intersects the dispersion surface at an angle $\theta$, giving a well defined group velocity of the electron wavepacket (Figure \ref{fig:Fig1}a). The intersection curve, called the spectral support curve, can be projected onto the $k_x-\omega$ plane (Figure \ref{fig:Fig1}b), which shares at least one point $(k_0,\omega_0)$ with the spectral plane. We call this lowest frequency $\omega_0$ the carrier frequency, as seen in the $k_z-\omega$ plane (Figure \ref{fig:Fig1}c). 
Tuning the parameters $k_{0}$ and $\theta$ (the tilt of the spectral plane) gives different spectral projections onto the plane $k_x-\omega$ and $k_z-\omega$ (Figure \ref{fig:Fig1}d,e,f). In this paper, we focus on the superluminal electrons with group velocities above the carrier group velocity at ($k_0,\omega_0$).

These projections form the energy-momentum relations that the electron plane waves should possess to form a single electron STWP. Upon introducing the dimension $k_y$, the curve in Figures \ref{fig:Fig1}b,e is revolved into a three-dimensional surface in momentum space. The different topologies of the three dimensional surfaces, either closed or open, is used to form momentum space skyrmions and fractional skyrmions respectively. This is due to the momentum space surface covering a full Bloch sphere, or a portion of the Bloch sphere. This will be detailed more below. We first give an analytic description of the range where momentum space merons and skyrmions can be hosted in an electron STWP based on the three dimensional topology of the spectral surface and for a given maximum electron energy with frequency $\omega_{max}$.

\begin{figure}[htbp]
    \centering
    \includegraphics[width=1\linewidth, height=9.5cm]{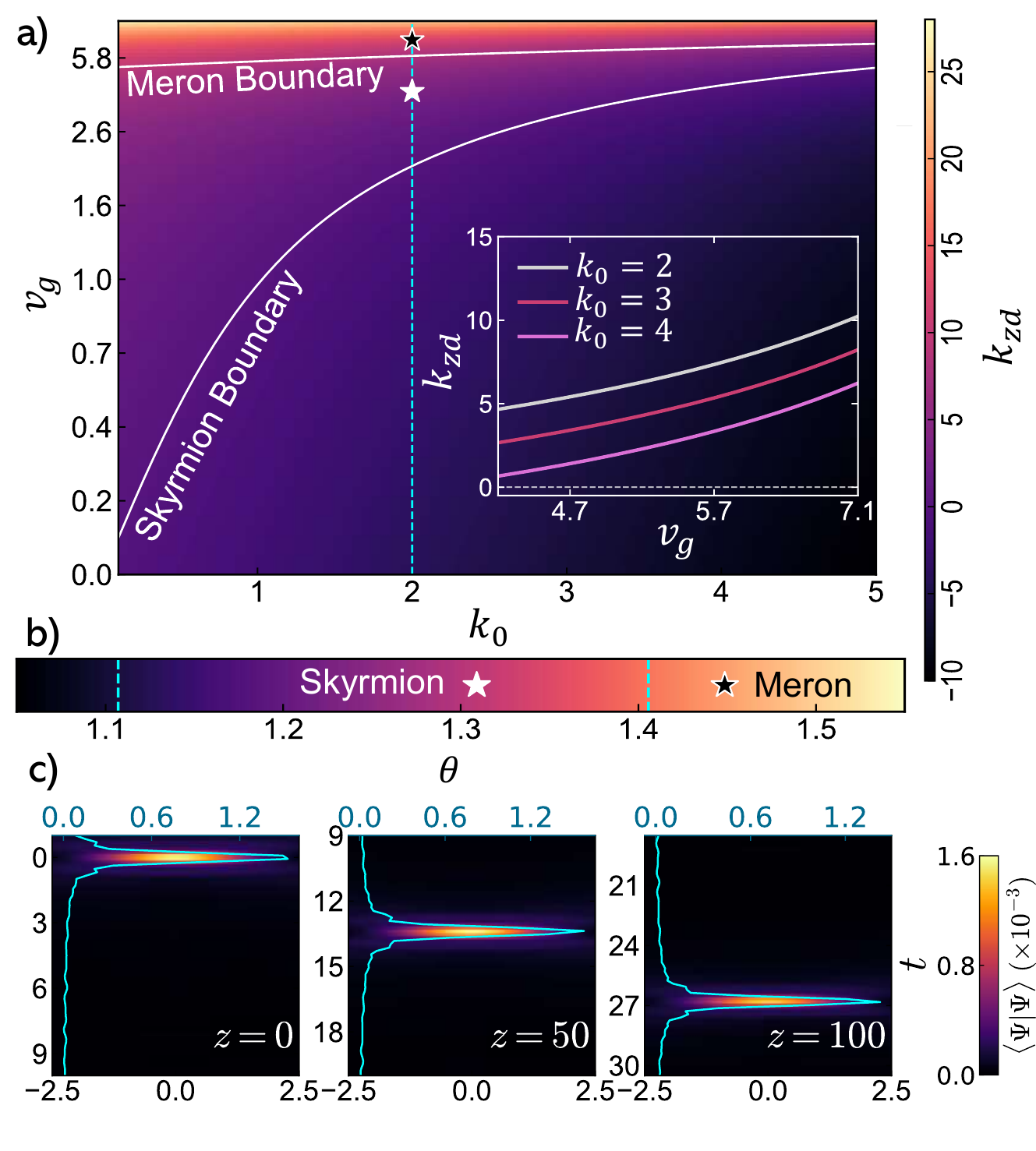} %fig-1.jpg
    \caption{a) The allowed parameter space in $v_g,k_0$ to host skyrmion and meron in momentum space in an electron STWP. The colormap is the longitudinal spectral linewidth in momentum space, and the white and black stars are the points used for skyrmion and meron real-space calculations respectively. Inset: The longitudinal spectral linewidth as a function of group velocity and carrier wavenumbers. The shown $k_{zd}$ is measured without a limit $k_{zmax}$. b) The group velocity regions for a carrier wavenumber of $k_0=2$, shown using the spectral tilt angle. c) The propagation invariance of the electron STWP can be seen by the almost unchanged peak intensity  (blue line and axis) as the pulse moves form $z=0$ to $z=100$.  The probability amplitude plot is scaled by $10^{3}$ for clarity, and shown in the lab frame. }
    \label{fig:Fig2}
\end{figure}

To investigate the topology, the value of the carrier wavenumber and the group velocity was used as free parameters. First, by imposing a condition where the spectral plane intersects the dispersion surface at exactly the carrier frequency, the lowest group velocity where the electron wavepackets has to possess to get a closed momentum space surface is obtained. This is equivalent to the gradient of the dispersion surface $\partial\omega / \partial k_z$ at the point ($k_0,\omega_0$). The group velocity has to satisfy Equation \ref{eq:thetas} for the electron wavepacket to host a momentum space skyrmion.
\begin{equation}
    v_g > \frac{\hbar k_0}{m}
    \label{eq:thetas}
\end{equation}
At the same time, there is an upper limit after which only open momentum space surfaces can be obtained, and therefore only merons can be hosted. The analytical starting point depends on a given maximum electron energy making up the wavepacket. We note that the largest group velocity that can host a skyrmion is when the spectral plane cuts the dispersion surface at exactly $k_{0}$ and $k_{zmax}$, where $k_{zmax}$ is the maximum value of the longitudinal wavenumber where the spectral plane can intersect the dispersion surface (Figure \ref{fig:Fig1}c and the corresponding figure caption, $k_{zmax}=10$ in Figure \ref{fig:Fig2}). This directly leads to the constraint in Equation \ref{eq:thetam}. Both group velocity constraints for merons and skyrmions are visualized in the Figure \ref{fig:Fig2}a. 
 \begin{equation}
    v_g > \frac{\hbar (k_{zmax}^2-k_0^2)}{2m(k_{zmax}-k_0)}
    \label{eq:thetam}
\end{equation}
Within the region that can host skyrmions, the closed three dimensional surface spanned by $(k_x,k_y,\omega)$ also has a size dependence on the group velocity and the carrier frequency. This size dependence is captured by the difference between the two values of $k_z$ where the spectral plane intersects the dispersion surface (Figure \ref{fig:Fig1}c). These points are given by $k_0$ and Equation \ref{eq:deltakz} (See Supplemental Material A).
\begin{equation}
    k_{z}' = \frac{2m \tan(\theta)}{\hbar} - k_0
    \label{eq:deltakz}
\end{equation}
With $\theta\geq \arctan(\hbar k_0 /m)$ so all the spectral components have $k_z >0 $, constrained to be only a forward propagating electron wave.

The colormap on Figure \ref{fig:Fig2}a depicts the different sizes of the momentum space surface for different values of the group velocity, $v_g$ and the carrier wavenumber, $k_0$. For higher values of $k_0$, the lower $v_g$ have a negative longitudinal spectral linewidth $k_{zd} = k_{z}'-k_0$, simply due to the fact that the point of intersection defined in the spectral plane, $k_0$ will be the maximum instead of the minimum. This is the region of subluminal electrons, where the group velocity falls below the group velocity at $(k_0,\omega_0)$ in Figure \ref{fig:Fig1}c. For higher $k_0$, it will also be increasingly difficult to construct a momentum space skyrmion before reaching the topological phase transition into a momentum space meron, despite higher energy electrons should also be treated relativistically.% as the group velocity drops below the Skyrmion boundary.
%and the spectral support curve may include negative $k_z$.

Within the regime where a skyrmion can be hosted, a higher $k_0$ with a fixed $v_g$ gives a lower size of the skyrmion in momentum space (Figure \ref{fig:Fig2}a inset), which directly impacts the spectral requirement of the electron plane waves required to synthesize the whole single electron wavepacket. In reality, this difference in the spectral extent in the synthesis of the STWP is directly observable on the probability distribution in real-space. Keeping $k_0 = 2.0$, the group velocity was tuned to construct the different momentum space surfaces and prove the boundaries given by the analytical derivations. The white star are the parameters used to construct the skyrmion and the black star are the parameters for the meron. Figure \ref{fig:Fig2}b shows the spectral tilt angle used instead of the group velocity for clarity. 

Constructing the electron wavefunction with the allowed ($k_x,k_y,\omega$) plane wave components on the momentum space surface is enough to create a propagation invariant and nondiffracting electron STWP (see Figure \ref{fig:Fig2}c, the blue line and axes are the intensity at $x=0$ overlaid onto the probability amplitude plot). However, such a scalar STWP construction lacks a vector structure on the momentum space surface. Therefore, we detail now how the momentum space skyrmions and merons are constructed. We use the nonrelativistic electron spinor and impose it onto each point on the surface in momentum space. Effectively, this forms a momentum space skyrmion (similar to a momentum space meron in \cite{Yessenov_Dorrah_Guo_Hall_Park_Free_Johnson_Capasso_Fan_Abouraddy_2025}) and in real-space, it forms a propagating localized electron wavepacket made up of momentum-spinor coupled plane waves. 

\begin{figure}
    \centering
    \includegraphics[width=1\linewidth,height=6.45cm]{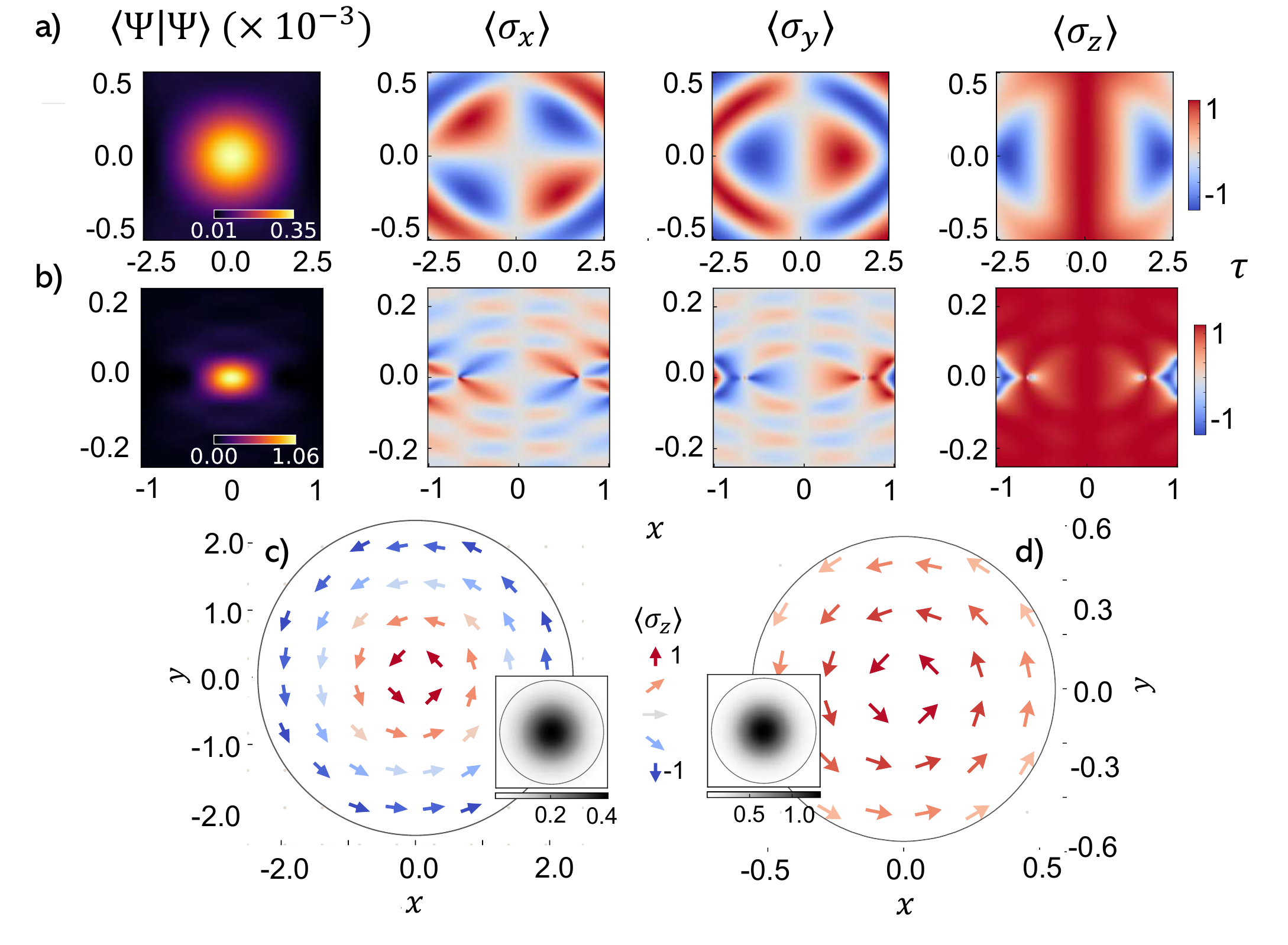} %fig--32.jpg
    \caption{ The realspace structure of the electron wavefunction. a,b) The spatiotemporal structure for skyrmions and meron SEV respectively in the comoving frame. The SEV is normalized by the local probability amplitude. c) The real-space SEV of the skyrmions and d) merons. The in-plane arrows are directions of $\langle \sigma_x \rangle$ and $\langle \sigma_y \rangle$, while the colormap and legend gives the direction of $\langle \sigma_z \rangle$. Inset: probability map of the electron pulse with the skyrmion and meron structure embedded, scaled to $10^3$.}
    \label{fig:Fig3}
\end{figure}

The spinor is defined as a two component orthogonal state as Eq.~(\ref{eq:e1}), which leads to the vectors on a Bloch sphere being defined as $\overline n = [\sin\theta \cos \phi, \;\sin\theta \sin \phi ,\;\cos\theta]$. Here $m,\gamma$ is the vorticity and the helicity of the topological texture. In this paper, $m,\gamma$ are set to 1 and 0 respectively. We adopt a polarity of 1 for the momentum space skyrmions and polarity of -1 for the momentum space merons (See Supplemental Material C). 
\begin{equation}
    \chi = [\cos(\theta/2), e^{i(m\phi+\gamma)}\sin(\theta/2)]
    \label{eq:e1}
\end{equation}
As the electron spin is an intrinsic quantity, we investigate the local spin expectation value (SEV) of the electron wavefunction in real-space, given by $\langle\Psi|\boldsymbol{\sigma}|\Psi\rangle$, with $\Psi$ being the real-space wavefunction $\Psi = \int A e^{i(k_xx+k_yy+k_zz-\omega t)}\chi(k_x,k_y) dk_xdk_y$. $\boldsymbol{\sigma}$ is the three component Pauli matrices, and A is the amplitude of each constituent plane wave from each point on the momentum space surface and $\Psi = (\Psi_\uparrow,\Psi_\downarrow)$ is a two-component normalized wavefunction corresponding to the components of $\chi$. The governing equation for $\Psi$ should ideally be the Pauli equation, but with the absence of magnetic field, this is a two component field satisfying independent Schrodinger equations. We define the vector ${\bf S} = (\langle \sigma_x \rangle,\langle \sigma_y \rangle,\langle \sigma_z \rangle)$, which are the local normalized SEV (Eq.~(\ref{eq:en})), and we note that this information is carried within each single electron wavefunction.
\begin{equation}
    {\bf S}=\frac{
    \Psi^{\dagger}\boldsymbol{\sigma}\Psi
    }{
    \Psi^{\dagger}\Psi
    } = \frac{1}{\rho
    }
    \begin{pmatrix}
    2\,\operatorname{Re}
    \!\left(
    \Psi_{\uparrow}^{*}\Psi_{\downarrow}
    \right)
    \\[6pt]
    2\,\operatorname{Im}
    \!\left(
    \Psi_{\uparrow}^{*}\Psi_{\downarrow}
    \right)
    \\[6pt]
    |\Psi_{\uparrow}|^2-|\Psi_{\downarrow}|^2
    \end{pmatrix}
    \label{eq:en}
\end{equation}

where $\rho = |\Psi_{\uparrow}|^2+|\Psi_{\downarrow}|^2$.

The construction of the skyrmions and merons were done in momentum space, while Equation \ref{eq:en} was used to investigate the real-space probability distribution. Turning now to the behaviour in real-space, we projected the expectation values of the three Pauli matrices onto the $x-\tau$ plane (Figures \ref{fig:Fig3}a,b for skyrmions and merons respectively). For skyrmion SEV, the radial transition from red to blue indicates a reversal of the out of plane component in $\langle\Psi|\sigma_z|\Psi\rangle$, and pole-to-pole coverage of the Bloch sphere. In contrast, the meron SEV do not exhibit pole-to-pole coverage.  
\begin{figure}
    \centering
    \includegraphics[width=1\linewidth]{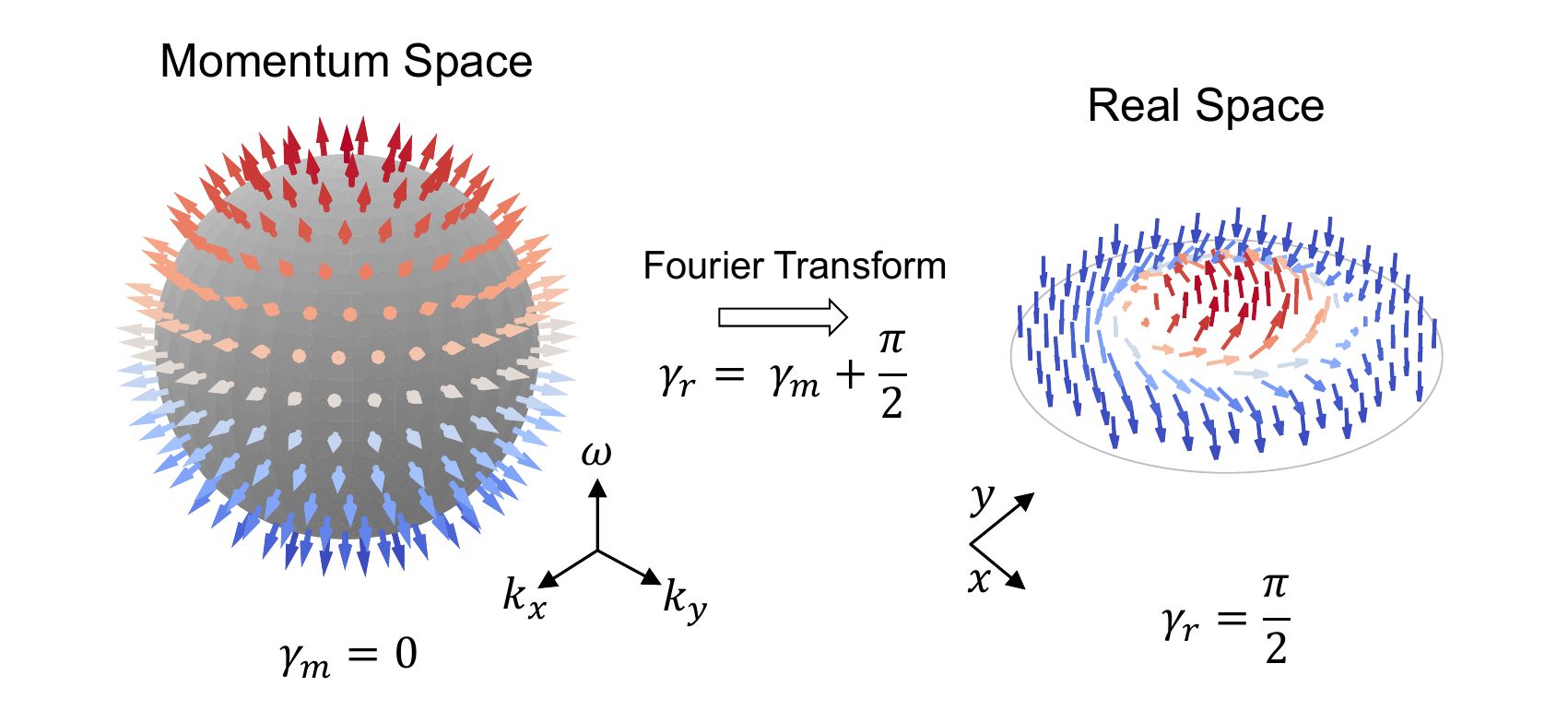} %kspacerspace.jpg
    \caption{The relation between the 3D Skyrmionic spinor texture in momentum space and its in-plane Fourier transformed counterpart in realspace. There is a helicity shift of $\pi/2$ due to an extra $i^m$ term, with $m=1$ being the winding number (Supplementary B). $\gamma_m$ is the helicity in momentum space of the spinor texture, and $\gamma_r$ is the realspace helicity.}
    \label{fig:kspace-realspace}
\end{figure}
Next, we plotted the SEV on the $x-y$ plane ($z,t=0)$ (Figure \ref{fig:Fig3}c,d for skyrmion and meron SEV respectively). For skyrmions, the center of the wavepacket is oriented purely in the positive $z$ direction ($\langle\sigma_z\rangle=1$, and the ends of the high probability region has the expectation value of $\langle\sigma_z\rangle=-1$. For merons, the ends of the high probability region instead gives $\langle\sigma_z\rangle=0$. Interestingly, the textures were found to exhibit a $\pi/2$ phase shift in the helicity (Figure \ref{fig:Fig3}c,d). A Néel skyrmion in momentum space gives a Bloch SEV in real-space, and the reason for this is an extra factor of $i$. Intuitively, the momentum space factor $e^{im\phi_k}$ is mapped to $i^me^{im\phi}$ in realspace. Hence, the integer winding $m$ and the topological character of the texture, through the skyrmion number, is preserved, but the additional phase $i^m$ rotates the in-plane spin orientation by $m\pi/2$. This accounts for the observed helicity shift for $m=1$ (Figure \ref{fig:kspace-realspace}). The Fourier transform therefore reshapes the radial profile through the Bessel function (See Supplemental Material B for further details). The existence of the skyrmion and meron local SEV permits the calculation of the skyrmion number in real-space (Equation \ref{eqn:sk-num}), which uses information encoded within each single electron wavefunction.
\begin{equation}
    Q = \frac{1}{4\pi}\int \int \mathbf{m}\cdot \left(\frac{\partial \mathbf{m}}{\partial x} \times \frac{\partial \mathbf{m}}{\partial y}\right) dx\; dy
    \label{eqn:sk-num}
\end{equation}
where $\mathbf{m}$ is the vector creating the texture. As the electron STWP adheres to the linear relation of $k_z$ and $\omega$ (Figure \ref{fig:Fig1}c,f), the matter wavepacket is propagation invariant for large distances. It is therefore of interest to measure the propagation invariance of the skyrmion number in realspace, and this is shown in Figure \ref{fig:Fig4} for skyrmions. The propagation invariance was measured in two ways. First, by changing the carrier frequency and wavenumber, and second, by tuning the group velocity. Figure \ref{fig:Fig4} shows both the measurements. Importantly, all the skyrmions propagate invariantly for large distances ($10^3$) before any large decays happen, which directly demonstrates that the tunable structures embedded within an electron STWP, are propagation invariant for relatively long distances. The difference between the different decays at large distances for different group velocities and carrier wavenumber is due to the finite spectral weights. An ideal STWP will show a perfectly flat profile across all distances. The Space-Time meron textures were already produced recently in optical STWP \cite{Yessenov_Dorrah_Guo_Hall_Park_Free_Johnson_Capasso_Fan_Abouraddy_2025}, which inherits the propagation invariance automatically from the STWP construction as well. 

\begin{figure}
    \centering
    \includegraphics[width=1\linewidth]{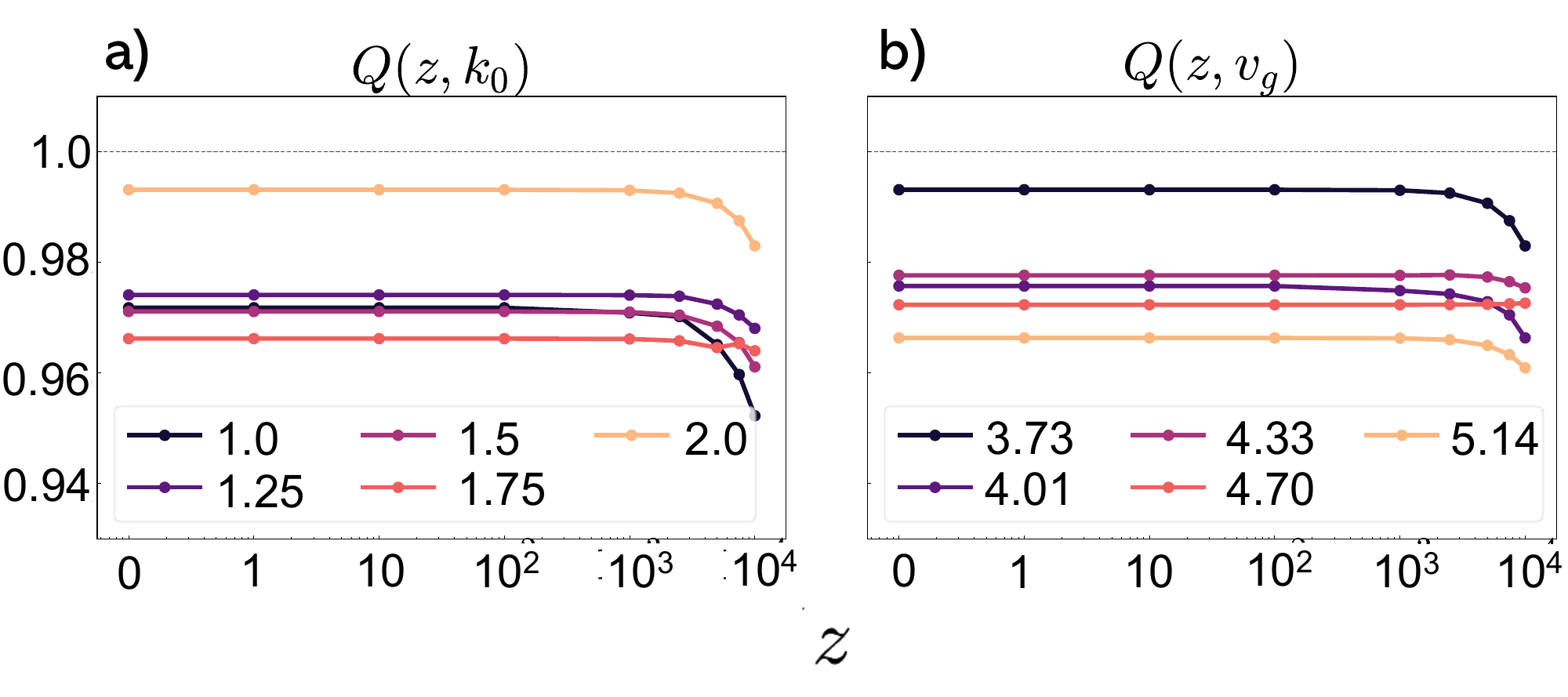} %fig-3.jpg
    \caption{ a) The propagation invariance of the skyrmion SEV for group velocity of $v_g = 3.73$ for different carrier wavenumbers, and b) the propagation invariance of the skyrmion SEV as a function of the group velocity, for $k_0=2.0$.}
    \label{fig:Fig4}
\end{figure}

In experiments, recent progress in electron waveshaping \cite{Madan_Leccese_Mazur_Barantani_LaGrange_Sapozhnik_Tengdin_Gargiulo_Rotunno_Olaya_etal._2022,Reinhardt_Kaminer_2020, McMorran_Agrawal_Anderson_Herzing_Lezec_McClelland_Unguris_2011,Lloyd_Babiker_Thirunavukkarasu_Yuan_2017,Fang_Kuttruff_Nabben_Baum_2024} may be used to sculpt the spatiotemporal spectrum of the electron wavepacket. To imprint spinor textures onto a free-electron plane waves in momentum space, a magnetic field may be utilized to give a probability distribution and realize the full or half Bloch sphere. 
This work is a general property of the quadratic dependence of the frequency on the momentum in the free electron dispersion relation. 
For other waves, the pseudospin degree of freedom intrinsic to quasiparticles may also be used \cite{He_Luo_Sohn_Matsuda_Wright_Su_2026, Herring_Kittel_1951}.

\textit{Conclusion - }In summary, we demonstrate that free space propagating STWPs can be created using nonrelativistic free-electron matter waves, and the electron spinor can be used to construct propagation invariant momentum space topological textures tunable by the group velocity and the carrier wavenumber. At high group velocities, the momentum space surface will only host merons, as there is no closed momentum space surface directly obtained through the intersection of the spectral plane and the dispersion surface. However, at lower group velocities, the parabolic nature of the free-electron dispersion naturally allows a closed sphere in ($k_x,k_y,\omega$) space. The closed momentum space structure allows the construction of momentum space skyrmions, and forms an analogous Bloch sphere. 

By this particular tuning, we show the possibility to generate persistent skyrmionic SEV structures in real-space that exhibits a helicity phase shift of $\pi/2$.

This paper reveals a new class of free-space wave packet with a topological wavefunction and charge, whose topology is independently tunable without material bandstructures. As this topological information exists in real-space, this may be a new robust route to transporting information with matter and for sensing applications.

Another property automatically inherited from the Space-Time Wave Packets structure is self healing \cite{Kondakci_Abouraddy_2018}, which allows the skyrmions and meron SEV proposed here to propagate for large distances even with small perturbations. This approach provides a bridge between STWP and structured matter waves, and the proposed system may also offer a platform for studying tunable topological textures in matter waves beyond conventional methods typically done through Bose-Einstein Condensates \cite{AlKhawaja_Stoof_2001,Ho_1998,Leslie_Hansen_Wright_Deutsch_Bigelow_2009,Marzlin_Zhang_Sanders_2000,Ruostekoski_Anglin_2001}. To the best of our knowledge, this is the first single electron wavepacket free space propagating skyrmion, shown via the local SEV of electrons, and we assert this is still a single electron state. Our work opens up a range of possibilities of integration with a large body of existing literature on optical skyrmions, and free-electron behaviour. Such electron skyrmionic wave packets may be relevant to free-electron quantum optics, free-electron-photon interactions, may provide additional degrees of freedom for free-electron entanglement, microscopy, free-electron-matter interaction and transition radiation and light generation by free-electrons \cite{Tsesses_Bartal_Kaminer_2017,Feranchuk_Ulyanenkov_2001,Konecna_Iyikanat_GarciaDeAbajo_2022,Ruimy_Karnieli_Kaminer_2025,Baranes_Ruimy_Gorlach_Kaminer_2022,Xie_Chen_Li_Yan_Chen_Lin_Kaminer_Miller_Yang_2025,Makarov_Makarova_Kapustin_Goshev_Tsykareva_2024}, as well as any future schemes for information encoding and transfer that will empower future technologies. 

\textbf{Acknowledgments.}
Singapore Ministry of Education (MOE) AcRF Tier 1 (RG157/23 \& RT11/23), Singapore Agency for Science, Technology and Research (A*STAR) (M24N7c0080 \& H25-MRO3489), and Nanyang Assistant Professorship Start Up grant. 

\textbf{Data Availability}
The numerical simulations that support the findings of this study are openly available at the following URL/DOI: \url{https://doi.org/10.21979/N9/EFUNNW} \cite{data_avail}

\bibliography{apssamp}

@article{SKYRME1962556,
title = {A unified field theory of mesons and baryons},
journal = {Nuclear Physics},
volume = {31},
pages = {556-569},
year = {1962},
issn = {0029-5582},
doi = {https://doi.org/10.1016/0029-5582(62)90775-7},
url = {https://www.sciencedirect.com/science/article/pii/0029558262907757},
author = {T.H.R. Skyrme}
}

@article{Psaroudaki_Panagopoulos_2021, title={Skyrmion Qubits: A New Class of Quantum Logic Elements Based on Nanoscale Magnetization}, volume={127}, ISSN={0031-9007, 1079-7114}, DOI={10.1103/PhysRevLett.127.067201}, number={6}, journal={Physical Review Letters}, author={Psaroudaki, Christina and Panagopoulos, Christos}, year={2021}, month=aug, pages={067201}}

@article{Shen_Wang_Fan_2025, title={Free-space topological optical textures: tutorial}, volume={17}, ISSN={1943-8206}, DOI={10.1364/AOP.547634}, abstractNote={Topological textures are well-established topics in condensed matter systems and nonlinear field theories. A typical example is the magnetic spin texture, which promises high-density data storage and information processing applications. With the recent development of nanophotonics and structured light, the topological optical textures, which are analogous to magnetic spin textures, can be created in linear electromagnetic fields with connections to solid-state physics but relying on radically different mechanisms. The emerging field of free-space topological optical textures has begun to show its ability to emulate diversified topologies in higher-dimensional light fields and open new directions of topologically protected information transfer. This article reviews the background of such topological textures, introduces a tutorial of fundamental theories for diverse topological textures in free space, and then provides perspective on the future potential applications to revolutionize our information society.}, number={2}, journal={Advances in Optics and Photonics}, author={Shen, Yijie and Wang, Haiwen and Fan, Shanhui}, year={2025}, month=jun, pages={295} }

@article{Shen_Zhang_Shi_Du_Yuan_Zayats_2024, title={Optical skyrmions and other topological quasiparticles of light}, volume={18}, ISSN={1749-4885, 1749-4893}, DOI={10.1038/s41566-023-01325-7}, number={1}, journal={Nature Photonics}, author={Shen, Yijie and Zhang, Qiang and Shi, Peng and Du, Luping and Yuan, Xiaocong and Zayats, Anatoly V.}, year={2024}, month=jan, pages={15–25}}

@article{Teng_Liu_Zhang_Fan_Chen_Cao_Zhong_Lei_Zhan_2025, title={Construction of optical spatiotemporal skyrmions}, volume={14}, ISSN={2047-7538}, DOI={10.1038/s41377-025-02028-0}, abstractNote={The creation and manipulation of photonic skyrmions provide a novel degree of freedom for light-matter interactions, optical communication and nanometrology. Since the localized vortex within skyrmions arises from the twist and curl of the phase structure, the orbital angular momentum of light is essential for their construction. While numerous skyrmionic textures have been proposed, they are formed within the spatial domain and induced by the longitudinal orbital angular momentum. Here we theoretically propose and experimentally observe spatiotemporal skyrmions within a picosecond pulse wavepacket, generated through vectorial sculpturing of spatiotemporal wavepackets. The skyrmionic textures emerge within the spatiotemporal distribution of a vector ﬁeld encompass all possible polarization states. Constructed upon the transverse orbital angular momentum, spatiotemporal skyrmions, in contrast to spatial skyrmions, exhibit no helical twisting perpendicular to the skyrmion plane, demonstrating potential stability against deformations or perturbations. These results expand the skyrmion family and offer new insights into optical quasiparticles, potentially leading to advanced applications in optical metrology, sensing, and data storage.}, number={1}, journal={Light: Science \& Applications}, author={Teng, Houan and Liu, Xin and Zhang, Nianjia and Fan, Haihao and Chen, Guoliang and Cao, Qian and Zhong, Jinzhan and Lei, Xinrui and Zhan, Qiwen}, year={2025}, month=sep, pages={324} }

@misc{data_avail,
  howpublished = {\url{https://doi.org/10.21979/N9/EFUNNW}},
}

@article{Zhan_2024, title={Spatiotemporal sculpturing of light: a tutorial}, volume={16}, ISSN={1943-8206}, DOI={10.1364/AOP.507558}, abstractNote={Spatiotemporal sculpturing of light pulses with sophisticated structures on demand is one major goal of the everlasting pursuit of ultrafast information transmission and processing as well as ultraintense energy concentration and extraction using light. It may hold the key to unlocking new extraordinary fundamental physical effects. Traditionally, spatiotemporal light pulses are treated as spatiotemporally separable wave packets as a solution to Maxwell’s equations. In the past decade, more generalized forms of spatiotemporally nonseparable solution started to emerge with growing importance for their striking physical effects. This tutorial intends to provide the necessary basics on how to sculpture light in the spatiotemporal domain to realize spatiotemporal structures on demand and highlight some of the recent advances in the creation and characterization of increasingly complex spatiotemporal wave packets. These spatiotemporally separable to complex nonseparable states with diverse geometric and topological structures exhibit unique physical properties during propagation, focusing, and interaction with matter. The broad potential applications as well as outlook and future trends and open challenges in this field are presented.}, number={2}, journal={Advances in Optics and Photonics}, author={Zhan, Qiwen}, year={2024}, month=jun, pages={163} }

@article{Guo_Peters_Mata-Cervera_Vetlugin_Guo_Zhang_Forbes_Shen_2026, title={Topological robustness of classical and quantum optical skyrmions in atmospheric turbulence}, volume={17}, ISSN={2041-1723}, DOI={10.1038/s41467-026-68751-3}, number={1}, journal={Nature Communications}, author={Guo, Zhenyu and Peters, Cade and Mata-Cervera, Nilo and Vetlugin, Anton N. and Guo, Ruixiang and Zhang, Pei and Forbes, Andrew and Shen, Yijie}, year={2026}, month=jan, pages={2085} }

@article{Wang_Zhao_Ma_Cai_Zhang_Shang_Zhang_Qin_Pong_Marozsak_et, title={Topological protection of optical skyrmions through complex media}, volume={13}, ISSN={2047-7538}, DOI={10.1038/s41377-024-01659-z}, abstractNote={Abstract
Optical Skyrmions have many important properties that make them ideal units for high-density data applications, including the ability to carry digital information through a discrete topological number and the independence of spatially varying polarization to other dimensions. More importantly, the topological nature of the optical Skyrmion heuristically suggests a strong degree of robustness to perturbations, which is crucial for reliably carrying information in noisy environments. However, the study of the topological robustness of optical Skyrmions is still in its infancy. Here, we quantify this robustness precisely by proving that the topological nature of the Skyrmion arises from its structure on the boundary and, by duality, is resilient to spatially varying perturbations provided they respect the relevant boundary conditions of the unperturbed Skyrmion. We then present experimental evidence validating this robustness in the context of paraxial Skyrmion beams against complex polarization aberrations. Our work provides a framework for handling various perturbations of Skyrmion fields and offers guarantees of robustness in a general sense. This, in turn, has implications for applications of the Skyrmion where their topological nature is exploited explicitly, and, in particular, provides an underpinning for the use of optical Skyrmions in communications and computing.}, number={1}, journal={Light: Science \& Applications}, author={Wang, An Aloysius and Zhao, Zimo and Ma, Yifei and Cai, Yuxi and Zhang, Runchen and Shang, Xiaoyi and Zhang, Yunqi and Qin, Ji and Pong, Zhi-Kai and Marozsák, Tádé and Chen, Binguo and He, Honghui and Luo, Lin and Booth, Martin J. and Elston, Steve J. and Morris, Stephen M. and He, Chao}, year={2024}, month=nov, pages={314}}

@article{Niu_Feng_Zhang_Yu_Wang_Lang_Xu_Chen_Ma_Qiu_et, title={Electric-magnetic-switchable free-space skyrmions in toroidal light pulses via a nonlinear metasurface}, volume={13}, ISSN={2334-2536}, DOI={10.1364/OPTICA.578501}, abstractNote={Recent advances reveal that light propagation in free space supports many exotic topological textures, such as skyrmions. Their unique space–time topologies make them promising candidates as next-generation robust information carriers. Hence, the ability of switching different texture modes to serve as a manner of data transfer is in high demand. However, previous studies focus on generation of one specific mode, lacking integrated devices with externally variable and stable mode generation capability. Here, we experimentally demonstrate the realization of switchable skyrmions between electric and magnetic modes in toroidal light pulses (TLPs) using a nonlinear metasurface platform in terms of broadband terahertz (THz) generation driven by a vectorial pulse. The spatial and temporal evolutions of them are also clearly observed. Our work establishes a new paradigm, to our knowledge, for manipulating and switching topologically structured light.}, number={2}, journal={Optica}, author={Niu, Li and Feng, Xi and Zhang, Xueqian and Yu, Wangke and Wang, Qingwei and Lang, Yuanhao and Xu, Quan and Chen, Xieyu and Ma, Jiajun and Qiu, Haidi and Shen, Yijie and Zhang, Weili and Han, Jiaguang}, year={2026}, month=feb, pages={203}}

@article{Shen_Hou_Papasimakis_Zheludev_2021, title={Supertoroidal light pulses as electromagnetic skyrmions propagating in free space}, volume={12}, ISSN={2041-1723}, DOI={10.1038/s41467-021-26037-w}, abstractNote={Abstract
            Topological complex transient electromagnetic fields give access to nontrivial light-matter interactions and provide additional degrees of freedom for information transfer. An important example of such electromagnetic excitations are space-time non-separable single-cycle pulses of toroidal topology, the exact solutions of Maxwell’s equations described by Hellwarth and Nouchi in 1996 and recently observed experimentally. Here we introduce an extended family of electromagnetic excitation, the supertoroidal electromagnetic pulses, in which the Hellwarth-Nouchi pulse is just the simplest member. The supertoroidal pulses exhibit skyrmionic structure of the electromagnetic fields, multiple singularities in the Poynting vector maps and fractal-like distributions of energy backflow. They are of interest for transient light-matter interactions, ultrafast optics, spectroscopy, and toroidal electrodynamics.}, number={1}, journal={Nature Communications}, author={Shen, Yijie and Hou, Yaonan and Papasimakis, Nikitas and Zheludev, Nikolay I.}, year={2021}, month=oct, pages={5891} }

@article{Wang_Bao_Hu_Shi_Wang_Zheludev_Shen_2024, title={Observation of resilient propagation and free-space skyrmions in toroidal electromagnetic pulses}, volume={11}, ISSN={1931-9401}, DOI={10.1063/5.0218207}, abstractNote={Toroidal electromagnetic pulses have been recently reported as nontransverse, space-time nonseparable topological excitations of free space. However, their propagation dynamics and topological configurations have not been comprehensively experimentally characterized. In addition, the existing generators were limited in optical and terahertz domains; however, the feasibility and significance of generating such pulses at microwave frequencies have been overlooked. Here, we report that microwave toroidal pulses can be launched by a transient finite-aperture broadband horn antenna emitter, as an electromagnetic counterpart of “air vortex cannon.” Applying this effective generator, we experimentally map the toroidal pulses’ topological skyrmionic textures in free space and demonstrate their resilient propagation dynamics, i.e., how that, during propagation, the pulses evolve toward stronger space-time nonseparability and closer proximity to the canonical Hellwarth–Nouchi toroidal pulses. Our work offers a practical opportunity for using topologically robust toroidal pulses as information carriers in high-capacity telecom, cell phone technology, remote sensing, and global positioning, especially where microwave frequencies are predominant.}, number={3}, journal={Applied Physics Reviews}, author={Wang, Ren and Bao, Pan-Yi and Hu, Zhi-Qiang and Shi, Shuai and Wang, Bing-Zhong and Zheludev, Nikolay I. and Shen, Yijie}, year={2024}, month=sep, pages={031411} }

@article{Wang_Ying_Shi_Wang_Wang_Liang_Shen_2025, title={Hybrid electromagnetic toroidal vortices}, volume={11}, ISSN={2375-2548}, DOI={10.1126/sciadv.ads4797}, abstractNote={The ubiquitous occurrence of toroidal vortices, or vortex rings, in fluid dynamics scenarios in nature has garnered notable attention in the scientific community, while their electromagnetic counterparts have only been proposed recently with two distinct manifestations: vector toroidal pulses and scalar phase toroidal vortices. Herein, we theoretically propose a form of electromagnetic toroidal vortex solutions that uniquely integrates both scalar and vector characteristics, challenging the prevailing notion of their mutual exclusivity. We also present an experimental generation of hybrid toroidal vortex pulses by a compact coaxial horn emitter augmented with a radial metasurface. These topological pulses exhibit peculiar electromagnetic features, such as vortex streets, skyrmions, and transverse orbital angular momentum, and may present advantages when propagating through perturbations, opening avenues for enhanced free-space information transmission, topologically nontrivial light-matter interaction, and microscopy techniques.
, Toroidal electromagnetic pulses are upgraded to host vortex streets, skyrmion topology, and transverse orbital angular momentum.}, number={8}, journal={Science Advances}, author={Wang, Ren and Ying, Beier and Shi, Shuai and Wang, Junsong and Wang, Bing-Zhong and Liang, Musheng and Shen, Yijie}, year={2025}, month=feb, pages={eads4797} }

@article{Shen_Papasimakis_Zheludev_2024, title={Nondiffracting supertoroidal pulses and optical “Kármán vortex streets”}, volume={15}, ISSN={2041-1723}, DOI={10.1038/s41467-024-48927-5}, abstractNote={Abstract
            Supertoroidal light pulses, as space-time nonseparable electromagnetic waves, exhibit unique topological properties including skyrmionic configurations, fractal-like singularities, and energy backflow in free space, which however do not survive upon propagation. Here, we introduce the non-diffracting supertoroidal pulses (NDSTPs) with propagation-robust skyrmionic and vortex field configurations that persists over arbitrary propagation distances. Intriguingly, the field structure of NDSTPs has a similarity with the von Kármán vortex street, a pattern of swirling vortices in fluid and gas dynamics with staggered singularities that can stably propagate forward. NDSTPs will be of interest as directed channels for information and energy transfer applications.}, number={1}, journal={Nature Communications}, author={Shen, Yijie and Papasimakis, Nikitas and Zheludev, Nikolay I.}, year={2024}, month=jun, pages={4863} }

@article{Zdagkas_McDonnell_Deng_Shen_Li_Ellenbogen_Papasimakis_Zheludev_2022, title={Observation of toroidal pulses of light}, volume={16}, ISSN={1749-4885, 1749-4893}, DOI={10.1038/s41566-022-01028-5}, number={7}, journal={Nature Photonics}, author={Zdagkas, Apostolos and McDonnell, Cormac and Deng, Junhong and Shen, Yijie and Li, Guixin and Ellenbogen, Tal and Papasimakis, Nikitas and Zheludev, Nikolay I.}, year={2022}, month=jul, pages={523–528} }

@article{Guo_Xiao_Guo_Yuan_Fan_2020, title={Meron Spin Textures in Momentum Space}, volume={124}, ISSN={0031-9007, 1079-7114}, DOI={10.1103/PhysRevLett.124.106103}, number={10}, journal={Physical Review Letters}, author={Guo, Cheng and Xiao, Meng and Guo, Yu and Yuan, Luqi and Fan, Shanhui}, year={2020}, month=mar, pages={106103} }

@article{Yessenov_Dorrah_Guo_Hall_Park_Free_Johnson_Capasso_Fan_Abouraddy_2025, title={Ultrafast space-time optical merons in momentum-energy space}, volume={16}, ISSN={2041-1723}, DOI={10.1038/s41467-025-63641-6}, number={1}, journal={Nature Communications}, author={Yessenov, Murat and Dorrah, Ahmed H. and Guo, Cheng and Hall, Layton A. and Park, Joon-Suh and Free, Justin and Johnson, Eric G. and Capasso, Federico and Fan, Shanhui and Abouraddy, Ayman F.}, year={2025}, month=sep, pages={8592}}

@article{Hall_Abouraddy_2025, title={Spectral reorganization of space-time wave packets in the presence of normal group-velocity dispersion}, volume={111}, ISSN={2469-9926, 2469-9934}, DOI={10.1103/ymls-2lkq}, number={6}, journal={Physical Review A}, author={Hall, Layton A. and Abouraddy, Ayman F.}, year={2025}, month=jun, pages={063515}}

@article{Kim_Guo_Catrysse_Fan_2025, title={Shaping space-time wave packets beyond the paraxial limit using a dispersion magnifier}, volume={111}, ISSN={2469-9926, 2469-9934}, DOI={10.1103/PhysRevA.111.013504}, number={1}, journal={Physical Review A}, author={Kim, Dongha and Guo, Cheng and Catrysse, Peter B. and Fan, Shanhui}, year={2025}, month=jan, pages={013504}}

@article{Kondakci_Abouraddy_2016, title={Diffraction-free pulsed optical beams via space-time correlations}, volume={24}, rights={https://doi.org/10.1364/OA_License_v1#VOR-OA}, ISSN={1094-4087}, DOI={10.1364/OE.24.028659}, number={25}, journal={Optics Express}, author={Kondakci, H. Esat and Abouraddy, Ayman F.}, year={2016}, month=dec, pages={28659}}

@article{Kondakci_Abouraddy_2019, title={Optical space-time wave packets having arbitrary group velocities in free space}, volume={10}, ISSN={2041-1723}, DOI={10.1038/s41467-019-08735-8}, abstractNote={Abstract
            Controlling the group velocity of an optical pulse typically requires traversing a material or structure whose dispersion is judiciously crafted. Alternatively, the group velocity can be modified in free space by spatially structuring the beam profile, but the realizable deviation from the speed of light in vacuum is small. Here we demonstrate precise and versatile control over the group velocity of a propagation-invariant optical wave packet in free space through sculpting its spatio-temporal spectrum. By jointly modulating the spatial and temporal degrees of freedom, arbitrary group velocities are unambiguously observed in free space above or below the speed of light in vacuum, whether in the forward direction propagating away from the source or even traveling backwards towards it.}, number={1}, journal={Nature Communications}, author={Kondakci, H. Esat and Abouraddy, Ayman F.}, year={2019}, month=feb, pages={929} }

@article{Romer_Hall_Abouraddy_2025, title={Synthesis and characterization of space-time light sheets: a tutorial}, volume={27}, ISSN={2040-8978, 2040-8986}, DOI={10.1088/2040-8986/ad93e5}, abstractNote={Abstract
            Space-time wave packets (STWPs) are a new class of pulsed optical beams with many unique and intriguing attributes, including propagation invariance and tunable group velocity in linear optical media. STWPs are a form of spatiotemporally structured light, so their synthesis poses challenges that are not shared by conventional monochromatic structured light fields. We present here a detailed description of the synthesis of STWPs that are localized along one transverse dimension and uniform along the other; i.e. space-time light sheets. We also describe the main characterization schemes needed for benchmarking the unique properties of space-time light sheets.}, number={1}, journal={Journal of Optics}, author={Romer, Miguel A and Hall, Layton A and Abouraddy, Ayman F}, year={2025}, month=jan, pages={013501} }

@article{Su_Zou_Wang_Yessenov_Zhou_Song_Ko_Zeng_Alhaddad_Ramakrishnan_et, title={Space-time wave packets in multimode optical fibers with controlled dynamic motions and tunable group velocities}, volume={16}, ISSN={2041-1723}, DOI={10.1038/s41467-025-56982-9}, number={1}, journal={Nature Communications}, author={Su, Xinzhou and Zou, Kaiheng and Wang, Yingning and Yessenov, Murat and Zhou, Huibin and Song, Hao and Ko, Wing and Zeng, Ruoyu and Alhaddad, Abdulrahman and Ramakrishnan, Murale and Duan, Yuxiang and Abouraddy, Ayman F. and Tur, Moshe and Christodoulides, Demetrios N. and Willner, Alan E.}, year={2025}, month=feb, pages={2027} }

@article{Wang_Guo_Fan_2025, title={Spatiotemporal Steering of Nondiffracting Wave Packets}, volume={134}, ISSN={0031-9007, 1079-7114}, DOI={10.1103/PhysRevLett.134.073803}, number={7}, journal={Physical Review Letters}, author={Wang, Haiwen and Guo, Cheng and Fan, Shanhui}, year={2025}, month=feb, pages={073803} }

@article{Yessenov_Abouraddy_2020, title={Accelerating and Decelerating Space-Time Optical Wave Packets in Free Space}, volume={125}, ISSN={0031-9007, 1079-7114}, DOI={10.1103/PhysRevLett.125.233901}, number={23}, journal={Physical Review Letters}, author={Yessenov, Murat and Abouraddy, Ayman F.}, year={2020}, month=dec, pages={233901} }

@article{Yessenov_Free_Chen_Johnson_Lavery_Alonso_Abouraddy_2022, title={Space-time wave packets localized in all dimensions}, volume={13}, ISSN={2041-1723}, DOI={10.1038/s41467-022-32240-0}, abstractNote={Abstract
            
              Optical wave packets that are localized in space and time, but nevertheless overcome diffraction and travel rigidly in free space, are a long sought-after field structure with applications ranging from microscopy and remote sensing, to nonlinear and quantum optics. However, synthesizing such wave packets requires introducing non-differentiable angular dispersion with high spectral precision in two transverse dimensions, a capability that has eluded optics to date. Here, we describe an experimental strategy capable of sculpting the spatio-temporal spectrum of a generic pulsed beam by introducing arbitrary radial chirp via two-dimensional conformal coordinate transformations of the spectrally resolved field. This procedure yields propagation-invariant ‘space-time’ wave packets localized in all dimensions, with tunable group velocity in the range from 0.7
              c
              to 1.8
              c
              in free space, and endowed with prescribed orbital angular momentum. By providing unprecedented flexibility in sculpting the three-dimensional structure of pulsed optical fields, our experimental strategy promises to be a versatile platform for the emerging enterprise of space-time optics.}, number={1}, journal={Nature Communications}, author={Yessenov, Murat and Free, Justin and Chen, Zhaozhong and Johnson, Eric G. and Lavery, Martin P. J. and Alonso, Miguel A. and Abouraddy, Ayman F.}, year={2022}, month=aug, pages={4573} }

@article{Yessenov_Hall_Schepler_Abouraddy_2022, title={Space-time wave packets}, volume={14}, ISSN={1943-8206}, DOI={10.1364/AOP.450016}, abstractNote={Space-time wave packets (STWPs) constitute a broad class of pulsed optical fields that are rigidly transported in linear media without diffraction or dispersion, and are therefore propagation-invariant in the absence of optical nonlinearities or waveguiding structures. Such wave packets exhibit unique characteristics, such as controllable group velocities in free space and exotic refractive phenomena. At the root of these behaviors is a fundamental feature underpinning STWPs: their spectra are not separable with respect to the spatial and temporal degrees of freedom. Indeed, the spatiotemporal structure is endowed with non-differentiable angular dispersion, in which each spatial frequency is associated with a single prescribed wavelength. Furthermore, controlled deviation from this particular spatiotemporal structure yields novel behaviors that depart from propagation-invariance in a precise manner, such as acceleration with an arbitrary axial distribution of the group velocity, tunable dispersion profiles, and Talbot effects in space–time. Although the basic concept of STWPs has been known since the 1980s, only very recently has rapid experimental development emerged. These advances are made possible by innovations in spatiotemporal Fourier synthesis, thereby opening a new frontier for structured light at the intersection of beam optics and ultrafast optics. Furthermore, a plethora of novel spatiotemporally structured optical fields (such as flying-focus wave packets, toroidal pulses, and spatiotemporal optical vortices) are now providing a swath of surprising characteristics, ranging from tunable group velocities to transverse orbital angular momentum. We review the historical development of STWPs, describe the new experimental approaches for their efficient synthesis, and enumerate the various new results and potential applications for STWPs and other spatiotemporally structured fields, before casting an eye on a future roadmap for this field.}, number={3}, journal={Advances in Optics and Photonics}, author={Yessenov, Murat and Hall, Layton A. and Schepler, Kenneth L. and Abouraddy, Ayman F.}, year={2022}, month=sep, pages={455}}

@article{Kondakci_Abouraddy_2017, title={Diffraction-free space–time light sheets}, volume={11}, ISSN={1749-4885, 1749-4893}, DOI={10.1038/s41566-017-0028-9}, number={11}, journal={Nature Photonics}, author={Kondakci, H. Esat and Abouraddy, Ayman F.}, year={2017}, month=nov, pages={733–740} }

@article{Che_Liu_Ye_Shi_Chan_Zi_2024, title={Generation of Spatiotemporal Vortex Pulses by Resonant Diffractive Grating}, volume={132}, ISSN={0031-9007, 1079-7114}, DOI={10.1103/PhysRevLett.132.044001}, number={4}, journal={Physical Review Letters}, author={Che, Zhiyuan and Liu, Wenzhe and Ye, Junyi and Shi, Lei and Chan, C. T. and Zi, Jian}, year={2024}, month=jan, pages={044001} }

@article{Liu_Ge_Xu_Sun_Liu_Lu_Chen_2025, title={Generation of spatiotemporal acoustic vortices with arbitrarily oriented orbital angular momentum}, volume={16}, ISSN={2041-1723}, DOI={10.1038/s41467-025-58154-1}, number={1}, journal={Nature Communications}, author={Liu, Shuai and Ge, Hao and Xu, Xiang-Yuan and Sun, Yuan and Liu, Xiao-Ping and Lu, Ming-Hui and Chen, Yan-Feng}, year={2025}, month=mar, pages={2823} }

@article{Singh_Ronning_Liu_Angheluta_Concha_Bandi_2026, title={Topology made visible through standing waves in a spinning fluid}, volume={9}, ISSN={2399-3650}, DOI={10.1038/s42005-026-02603-w}, number={1}, journal={Communications Physics}, author={Singh, Aditya and Rønning, Jonas and Liu, Chien-Chia and Angheluta, Luiza and Concha, Andres and Bandi, Mahesh M.}, year={2026}, month=apr, pages={123}}

@article{Smirnova_Nori_Bliokh_2024, title={Water-Wave Vortices and Skyrmions}, volume={132}, ISSN={0031-9007, 1079-7114}, DOI={10.1103/PhysRevLett.132.054003}, number={5}, journal={Physical Review Letters}, author={Smirnova, Daria A. and Nori, Franco and Bliokh, Konstantin Y.}, year={2024}, month=jan, pages={054003}}

@article{Baranes_Ruimy_Gorlach_Kaminer_2022, title={Free electrons can induce entanglement between photons}, volume={8}, DOI={10.1038/s41534-022-00540-4}, number={1}, journal={npj Quantum Information}, author={Baranes, Gefen and Ruimy, Ron and Gorlach, Alexey and Kaminer, Ido}, year={2022}, month={Mar}}

\end{document}